# INTRODUCTION TO PAPERS ON ASTROSTATISTICS

By Thomas J. Loredo, John Rice and Michael L. Stein

*Cornell University, University of California, Berkeley and University of Chicago*

We are pleased to present a Special Section on Statistics and Astronomy in this issue of the *The Annals of Applied Statistics*. Astronomy is an observational rather than experimental science; as a result, astronomical data sets both small and large present particularly challenging problems to analysts who must make the best of whatever the sky offers their instruments. The resulting statistical problems have enormous diversity. In one problem, one may have to carefully quantify uncertainty in a hard-won, sparse data set; in another, the sheer volume of data may forbid a formally optimal analysis, requiring judicious balancing of model sophistication, approximations, and clever algorithms. Often the data bear a complex relationship to the underlying phenomenon producing them, much in the manner of inverse problems.

There is a long history of fruitful interaction between astronomy and statistics; indeed, problems in astronomy and geodesy motivated many developments marking the emergence of the discipline of statistics in the early nineteenth century. Interaction between these disciplines has waxed and waned in the subsequent two centuries. Happily, the turn of the twenty-first century marks an era of renewed enthusiasm for collaborations between astronomers and statisticians, giving rise to the interdisciplinary area of astrostatistics. Six papers in this issue describe compelling problems in astrostatistics.

The main driver for renewed interaction between statistics and astronomy has been the advent of large-scale astronomical surveys. Modern surveys probe scales of space and time ranging from our own solar system, to planets in other solar systems, to the structure of the Milky Way and distant galaxies, to the very early history of the universe, using instruments that are sensitive to regions of the electromagnetic spectrum ranging from radio (wavelengths as long as 100 km) to gamma-rays (wavelengths as short as $10^{-6}$ nanometers). Observations using non-electromagnetic "messengers," such as cosmic ray particles, neutrinos, and gravitational waves, are also playing increasingly important roles in modern astronomy. The papers in the Special Section all address problems arising in the context of surveys.

For decades, astronomers have known that the motions of stars and galaxies indicate the presence of significant amounts of gravitating matter that does not emit observable light; there is quite literally more to the universe than meets the eye. Uncovering the nature of the so-called *dark matter* has long been one of the outstanding problems of astrophysics, particularly of cosmology, the study of the large-scale properties of the universe. A variety of cosmological observations in the last decade have greatly magnified the mystery of the universe's dark sector. New data finally allow precise measurement of the amount of dark matter: it is about five times more abundant than normal "baryonic" matter. Moreover, these observations also strongly indicate the existence of a strange and poorly understood *dark energy* field filling space, with the peculiar property that its net gravitational effect is to repel rather than to pull. Dark energy comprises about 75% of all the matter and energy in the universe, and it is presently accelerating the expansion of the universe. The sobering conclusion of this work is that about 95% of the universe consists of "stuff" utterly foreign to earthbound laboratories. Four of the papers here address diverse data analysis problems in cosmology, directly or indirectly associated with measuring the universe's dark sector.

One of the most significant sources of information about the early history of the universe on cosmological scales is the *cosmic microwave background* (CMB), a relic of the Big Bang, the hot and dense early phase of cosmic history. It is isotropic to about one part in 100,000, and instruments are measuring fluctuations with increasing sensitivity and on increasingly smaller angular scales; new instruments are measuring the even smaller polarization fluctuations of the CMB. These fluctuations mark the presence of slight density enhancements in the early universe that eventually became stars and galaxies (and scientists); their detailed properties tell us about cosmological initial conditions, the formation of cosmic structure, and the matter and energy content of the universe on large scales. Cabella and Marinucci review statistical challenges that arise in analysis of data from the *Wilkinson Microwave Anisotropy Probe* (*WMAP*), a NASA satellite mission providing the best all-sky measurements of the CMB. The challenges include estimation of angular power spectra and searching for non-Gaussianity and anisotropy with higher order spectra and certain specially constructed spherical wavelets.

The dark sector is detectable due to its gravitational effects. One of the most striking of these is gravitational lensing: the bending of the paths of light rays by gravitating matter or energy. In extreme cases, the bending can be so strong that concentrations of matter act as strong lenses, producing multiple images of distant objects. Much more commonly, matter concentrations produce *weak lensing*; rather than producing multiple images, dark matter concentrations slightly and systematically distort the shapes of distant galaxy images. Upcoming large-scale surveys will produce catalogs of galaxy images of unprecedented size, hopefully enabling use of weak lensing as a precise cosmological probe. Measurements of the shapes of many galaxies must be combined to yield information on the dark matter; the analysis must account for the diversity of galaxy shapes, and instrumental distortions. Devising statistical methods for this inverse problem is presented as a marvelous blinded challenge in the paper of Bridle et al. Will any readers seek out collaborators in astronomy and take up the gauntlet?

Dwarf spheroidal galaxies are small, ball-like concentrations of stars observed as satellites of the Milky Way and other large, nearby galaxies. Spectroscopic surveys of the motions of stars in these galaxies reveal that they are dominated by dark matter. Statisticians and astronomers have recently collaborated to develop nonparametric statistical methods to infer the dark matter distribution from the survey data, taking advantage of key prior information (monotonicity) to improve power. As part of this effort, the paper by Sen et al. models gravitational effects of the Milky Way on a particular galaxy with particular attention to streaming motion of leading and trailing stars as they move away from the center of the main body of the gravitationally perturbed system. Heavy use is made of nonparametric and bootstrap methods.

Dark energy was discovered in 1998 based on observations of Type Ia supernova (SN Ia) explosions, extremely luminous stellar-sized thermonuclear explosions visible to distances of billions of light years. These events have the unusual and valuable property of being a kind of "standard bomb." By observing the months-long unfolding of an explosion, astronomers can infer how intrinsically luminous it is; using a version of the inverse- square law, they can then infer how far away it is. Combining this with measurements of the explosion's host galaxy, astronomers can map how the expansion rate of the universe has evolved with time. The 1998 observations indicated the expansion rate is *increasing*, implying the presence of dark energy (other controversial explanations are also being explored). Effort has now turned to measuring how the basic properties of dark energy—its so-called "equation of state"—may be evolving in cosmic time; this will inform theories and future observations. Genovese et al. provide an eminently readable introduction to the area, and treat both parametric and nonparametric estimation and testing of hypotheses for the dark energy equation of state with new, optimal methods that address several weaknesses of many existing analyses.

The newest large space-based telescope mission is NASA's *Fermi Gamma-Ray Space Telescope* (formerly the *Gamma-ray Large Area Space Telescope*,

*GLAST*), launched in June 2008. *Fermi*'s task is to survey the entire sky in gamma-rays, the highest energy form of light. As we go to press, *Fermi* is about to announce its first results. Gamma-rays are so rare that astronomers must use instruments that can count individual gamma-ray quanta (photons). One phenomenon motivating the *Fermi* survey is high-energy pulsations from *pulsars*, neutron star remnants of the deaths of massive stars that spin with high regularity, producing periodic signals. Over a thousand pulsars are known from radio observations, but only a few are known to pulse in gamma-rays (or their lower-energy cousins, X-rays). Earlier gamma-ray surveys have found dozens of star-like sources that do not shine in other wavelength ranges; it is thought that many of these may be pulsars pulsing only in gamma-rays. *Fermi* will be able to detect faint pulsations, but detection requires "blind searching" for pulsations at an unknown frequency (and likely with significant linear frequency drift). The time series (photon arrival times) are sparse but very long; a naive exhaustive search is impossible. Meinshausen et al. describe a new adaptive blind search algorithm, first searching with a coarsened algorithm, then refining the search in promising regions. They specifically address *Fermi*'s pulsar searching, but formulate the problem more generally in terms of balancing statistical power within a recursive family of procedures against computational cost, using dynamic programming. This innovative approach surely has applications far beyond pulsar searching.

Finally, the mainstay of astronomy is, of course, visible stars, and the paper by van Dyk et al. addresses modeling the stellar population in a cluster of stars based on survey data comprised of luminosities and colors. As stars evolve, so do their luminosities and colors, and quantifications of the relationships of these two variables have long been used by astronomers to study stellar evolution. The analysis by van Dyk et al. brings a very natural Bayesian analysis to bear on combining observations with complex computer models of stellar evolution. Readers should find it interesting and instructive to see how contemporary Bayesian methods play out in this scenario.

These papers illustrate only a few of the opportunities in astrostatistics. One area that is missing is *data reduction*, the relatively low-level analysis of "raw" data that produces the data products listed in survey catalogs. For example, the process of photometry, whereby counts of photons in pixels of a CCD array are used to detect sources and measure their brightnesses and locations, presents challenges in statistical image processing, for example, due to imperfect instrumental response, atmospheric distortions, smooth but non-uniform backgrounds, and "source crowding" in dense fields. As another example, the process by which the angle of incidence of an incoming high energy photon in the X-ray or gamma-ray range is estimated using complex instrumentation is an interesting subject in itself. Astronomy is also a rich source of complex time series problems. Examples in which statistical techniques play an important role include the detection of extra-solar planets via motion or partial occultation of a host star, the detection and identification of transient phenomena such as gamma-ray bursts and supernovae, detection of gravitational waves (a nonelectromagnetic phenomenon), and the analysis of variable stars. For discussion of these and other topics, we recommend to the readers the proceedings volumes for the *Statistical Challenges in Modern Astronomy* conferences, edited by Jogesh Babu and Eric Feigelson, who have been promoting and developing astrostatistics for many years.

Contemporary astronomical data sets are already huge. For example, the Sloan Digital Sky Survey has cataloged nearly a billion galaxies (with detailed data for about 100 million), using a camera composed of 30 CCD chips each with a resolution of $2048 \times 2048$ pixels. With advancements in instrumentation and information technology, future surveys will produce data sets of almost mind-boggling enormity. For example, the Large Synoptic Survey Telescope (LSST), expected to begin operation in 2015, will produce around 30 terabytes of data per night from its three billion pixel camera, as it monitors everything from near-Earth asteroids to billions of remote galaxies. How to effectively deal with such enormous datasets presents fundamental challenges to the discipline of statistics in the twenty-first century.

T. J. Loredo
Department of Astronomy
Cornell University
610 Space Sciences Building
Ithaca, New York 14853
USA

J. Rice
Department of Statistics
University of California,
  Berkeley
367 Evans Hall
Berkeley, California 94720-3860
USA
E-mail: rice@stat.berkeley.edu

M. L. Stein
Department of Statistics
University of Chicago
5734 South University Avenue
Chicago, Illinois 60637
USA